\documentclass{revtex4}

\usepackage{ulem}
\usepackage{amsmath}

\begin{document}

\normalem

\title{Isoscalar \emph{g} Factors of Even-Even and Odd-Odd N=Z Nuclei}
\author{S. Yeager}
\affiliation{Department of Physics and Astronomy, Rutgers University
Piscataway, New Jersey 08854}
\author{L. Zamick}
\affiliation{Department of Physics and Astronomy, Rutgers University
Piscataway, New Jersey 08854}
\author{Y. Y. Sharon}
\affiliation{Department of Physics and Astronomy, Rutgers University
Piscataway, New Jersey 08854}
\author{S. J. Q. Robinson}
\affiliation{Department of Physics, Millsaps College
Jackson, Mississippi 39210}

\date{December 28, 2008}

\begin{abstract}
The $g$ factors of $2_1^+$ states in many even-even $N=Z$ nuclei, and also of $1^+$, $3^+$, and $5^+$ states in odd-odd $N=Z$ nuclei, have been measured to have values close to 0.5. Up-to-date compilations of the experimental values are presented. Their $g$ factors are all isoscalar. Although the collective $g$ factor for an $N=Z$ nucleus has a value of 0.5, we note that the Schmidt value in the large $l$ limit in a single $j$ shell also approaches 0.5, so care must also be taken in interpreting the results. When these expressions are evaluated, they yield $g$ factors whose values are surprisingly close to one half, which happens also to be the value of the collective $g$ factor for the states of $K=0$ bands in even-even $N=Z$ nuclei.  We also discuss briefly the $g$ factors of the ground states of mirror-nuclei pairs with closed $LS$ and $jj$ shells plus or minus one nucleon.
\end{abstract}

\maketitle

\section{Introduction}

In this work we focus on isoscalar magnetic moments. We consider excited states of $N=Z$ even-even nuclei,mainly $J = 2^+$ states, and also ground and excited
states of $N=Z$ odd-odd nuclei. To complete the picure we remind the reader of
previous works on mirror pairs of odd $A$ nuclei.

Isoscalar magnetic moments are much closer to the Schmidt values than the isovector ones. Nevertheless, there are small but systematic deviations. It was noted by Talmi \cite{talmi71} that
"The experimental values of $<$S$>$ seems to follow a simple rule. They are always smaller in absolute value than the values calculated in $jj$ coupling."

Arima, however, noted \cite{arima78} that the smallness of the isoscalar deviation
is due to the small isoscalar spin coupling (0.44) relative to that of the isovector coupling (2.353). If one divides the deviation by the lowest order result  one can get
a rather large ratio even in the isoscalar case,even up to 50\%.

In his monograph "Theories of Nuclear Moments" Blin-Stoyle \cite{blinstoyle57} discussed magnetic moments of odd $A$ and odd-odd nuclei. He did not discuss excited states of even-even nuclei since there was very little, if any, data available at the time.  The $jj$ value of the isoscalar moment for a single $j$ shell for a state of total angular momentum $I$ is

\begin{equation}
\begin{gathered}
\mu = gI \\
g = \frac{[g_j(p) + g_j(n)]}{2}
\end{gathered}
\end{equation}

where the $g$ factors are obtained from Schmidt magnetic moments ($\mu = gj$).

\begin{center}
\begin{tabular}{l l}
Odd Neutron & $\quad j = l+\dfrac{1}{2}$, $\qquad \mu _s$ = $\mu (n)$ \\
& $\quad j=l-\dfrac{1}{2}$, $\qquad \mu _s = \dfrac{-j}{j+1} \mu (n)$ \\
Odd Proton & $\quad j = l+\dfrac{1}{2}$, $\qquad \mu _s$ = $(j-\dfrac{1}{2}) + \mu (p)$ \\
& $\quad j = l-\dfrac{1}{2}$, $\qquad \mu _s$ = $\dfrac{j}{j+1}[(j+\dfrac{3}{2}) - \mu (p)]$
\end{tabular}
\end{center}

The formulae for isoscalar $g$ factors in the $LS$ limit, as listed by
Blin-Stoyle\cite{blinstoyle57}, are as follows

\begin{equation}
g_L = \frac{1}{2} + \frac{l_p(l_p +1)- l_n(l_n +1)}{2L(L+1)}
\end{equation}

\begin{center}
\begin{tabular}{ll  l}
&$I=L$, $S=0$ & $\qquad \mu = g_LI$ \\
&$I=L$, $S=1$ & $\qquad \mu = g_LI + \frac{(g_S - g_L)}{I+1}$ \\
&$I=L+1$, $S=1$ & $\qquad \mu = g_LI + g_S - g_L$ \\
&$I=L-1$, $S=1$ & $\qquad \mu = g_LI - \frac{(g_S - g_L)}{I+1}$ \\
\end{tabular}
\end{center}

The magnetic moments of light nuclei in both $jj$ and $LS$ coupling have been given by Talmi \cite{talmi51} and Blin-Stoyle \cite{blinstoyle57}. We will give a brief rediscussion in terms of $g$ factors, since $g$ factor systematics is the point of this work.

For the $2^+$ state of an even-even nucleus, the pure $LS$ configuration $S=0$ $J=L$ will yield a $g$ factor of 0.5. This point was emphasized by Arima \cite{arima78} to expain the closeness  of "$g=0.5$" for  $2^+$ states of $^{20}$Ne, $^{24}$Mg, $^{28}$Si, and $^{32}$S. We certainly agree with this but we will later contend that for heavier nuclei one must also consider the $jj$  limit systematics.

We note also that none of the $LS$ cases yield a value $g=0.5$ for odd-odd
nuclei. Here are some selected results

\begin{center} Table I - Selected odd-odd coupling results\\
\begin{tabular}{| c | c | c | c | c |} \hline
Nucleus & State & $LS$ & $jj$ & exp \\ \hline
$^2$H & $J=1^+$ & 0.88 & 0.88 & 0.88 \\ \hline
$^6$Li & $J=1^+$ & 0.88 & 0.627 & 0.8221 \\ \hline
$^{10}$B & $J=3^+$ & 0.627 & 0.627 & 0.600 \\ \hline
$^{14}$N & $J=1^+$ & 0.31 & 0.394 & 0.403 \\ \hline
\end{tabular}\end{center}

For $^6$Li, the $LS$ result for $L=0$ $S=1$ is the same as that for a deuteron. In $^{10}$B, the $J=3^+$ state has a unique configuration - the same in $LS$ and $jj$ - hence the same $g$ factor (which is also the same as the $g$ facor for $^6$Li in the $jj$ limit).

$^{14}$N is an interesting case. The $LS$ result is for $L=2$ $S=1$ which is closer to the true wave function than the $jj$ result $p_{1/2} (n)$ $p_{1/2} (p)$. It turns out there is a conspiracy of the spin orbit and tensor interactions that results in an almost good $LS$ configuration. This is at first surprising because in general the spin orbit force works against good $LS$ wave functions.

\section{The Experimental Data}

Table I gives the measured $g$ factors of the $2_1^+, \quad T=0$ states in even-even $N=Z$ nuclei. It also encompasses the $4_1^+, \quad T=0$ state in $^{20}$Ne.  In this connection, the table of N. Stone \cite{stone05} was very useful and the $^{36}$Ar data was from \cite{speidel06}. Also included in Table I (right hand column) are results of \uline{large}-space shell model calculations; the $^{32}$S and $^{36}$Ar calculated results are from \cite{speidel06}, the $^{44}$Ti calculated results are from \cite{schielke03}, and the others were computed by us (in excellent agreement with the earlier results of B.A. Brown \cite{brown82} and can be compared with the more accurate data now available. The experimental results \cite{speidel06, schielke03, leske03} for $^{20}$Ne, $^{32}$S, $^{36}$Ar, $^{44}$Ti were carried out by groups - those of Speidel and Koller - with whom the authors of the present paper have collaborated.

\begin{center}Table II - $g$ factors of 2$^+$, $T=0$ states \\
in even-even $N=Z$ nuclei \\
\begin{tabular}{| c | c | c |} \hline
Nucleus & Experiment & Calculated \\ \hline
$^{20}$Ne & 0.504(4)& 0.510 \\ \hline
$^{24}$Mg & 0.501(2)& 0.512 \\ \hline
$^{28}$Si & 0.55(10)& 0.514 \\ \hline
$^{32}$S & 0.44(10)&0.501 \\ \hline
& 0.47(9)& \\ \hline
$^{36}$Ar & 0.52(18)&0.488 \\ \hline
$^{44}$Ti & 0.52(15)&0.514 \\ \hline
$^{20}$Ne J=4 & 0.38(8)& 0.512 \\ \hline
\end{tabular}\end{center}

Table II gives the experimental data for the $g$ factors of $T=0$ states in odd-odd $N=Z$ nuclei.  The states considered have a total angular momentum of $J = 1^+$, $3^+$, $\textrm{or } 5^+$. Some of these states are ground states, so for them there is much greater precision.  The result for the $T=0$ ground state of $^{58}$Cu was obtained in 2008 by Stone et al. \cite{stone08}.

\begin{center}
\begin{tabular}{| c | c | c |} 
\multicolumn{3}{c}{Table III - Measured $g$ factors of}\\
\multicolumn{3}{c}{$T=0$ states in odd-odd $N=Z$ nuclei} \\ \hline
Nuclei & J & \emph{g} \\ \hline
$^2$H & 1$^+$ & 0.857438228(9) \\ \hline
$^6$Li & 1$^+$ & 0.8220473(6) \\ \hline
$^{10}$B & 3$^+$ & 0.60021493(2) \\ \hline
& 1$^+$ & 0.63(12) \\ \hline
$^{14}$N & 1$^+$ & 0.40376100(6) \\ \hline
$^{18}$F & 3$^+$ & 0.59(4) \\ \hline
& 5$^+$ & 0.572(6) \\ \hline
$^{22}$Na & 3$^+$ & 0.582(1) \\ \hline
& 1$^+$ & 0.523(11) \\ \hline
$^{26}$Al & 5$^+$ & 0.561(8) \\ \hline
$^{38}$K & 3$^+$ & 0.457(2) \\ \hline
$^{46}$V & 3$^+$ & 0.55(1) \\ \hline
$^{58}$Cu & 1$^+$ & 0.52(8) \\ \hline
\end{tabular} \end{center}

It can be noted from Tables II and III that except for the very light nuclei (e.g. the deuteron) the measured $g$ factors for these $T=0$ states in $N=Z$ nuclei have values that are not far away from 0.5, and that this is so for the odd-odd as well as the even-even $N=Z$ nuclei. This is the collective $g$ factor value in $N=Z$ nuclei for the states of a rotational $K=0$ band, $g = Z/A = 0.5$. However, it will be demonstrated below that a similar result of $g \cong 0.5$ is also predicted for these states by the single-$j$ shell model.

\section{$g$ Factors in the Single $j$ Shell}

It will be shown that in the single $j$ shell model the more complicated general $g$ factor expressions reduce to very simple ones for the $g$ factors in N=Z nuclei, both even-even and odd-odd. This expression is given in deShalit and Talmi \cite{deshalit63}. We will use expressions by McCullen et al. \cite{mccullen64} for $N \neq Z$ as a counterpoint to show how things simplify for $N=Z$.

More generally,  consider \uline{any} nucleus for which all the active protons and neutrons are in the same single $j$ shell. This would be the situation, for example, for the titanium isotopes with a closed $^{40}_{20}$Ca$_{20}$ core, where all the valence nucleons are in the $f_{7/2}$ shell with $j = 7/2$ and $l = 3$. The general expression for the $g$ factor of a state of total angular momentum $I$ for any Ti isotope is

\begin{equation}
\emph{g} = \frac{\emph{g}_{\emph{j}{P}} +\emph{g}_{\emph{j}{N}}}{2} + \frac{\emph{g}_{\emph{j}{P}} - \emph{g}_{\emph{j}{N}}}{2}\sum_{J_{P},J_{N}}|D^I(J_P,J_N)|^2 \frac{[J_P(J_P+1)-J_N(J_N+1)]}{I(I+1)}
\end{equation}

Here, $g_{j_{P}}$ and $g_{j_{N}}$ are the proton and neutron Schmidt $g$ factors in the $j$ shell under consideration which has the quantum numbers $(n$, $l$, $j)$. The second term is somewhat complicated \cite{mccullen64}, involving the factor $D^I(J_P, J_N)^2$ which is the probability that, in a state of total angular momentum I, the protons couple to an angular momentum $J_P$ and the neutrons to J$_N$. 

However, for an N=Z nucleus, due to charge symmetry, $D^I(J_P, J_N) = \pm D^I(J_N, J_P)$; this is true regardless of the isospin of the states. The sign does not matter because only the square enters into the expression (1). These facts, plus the presence of the factor $[J_{P}(J_{P}+1)-J_{N}(J_{N}+1)]$, make the second term drop out, and one is left with a very simple expression (given below) where the $g_j$, as noted above, are the Schmidt $g$ factors for the single $j$ shell under consideration.

\begin{equation}
g = \left(\frac{g_{j_{P}} + g_{j_{N}}}{2}\right)
\end{equation}

A big difference of the $LS$ and $jj$ limits for $N=Z$ nuclei is that in the $jj$ case, all states will have the same calculated $g$ factor as given by equation 4. This is certainly not true in the $LS$ limit.

Equation 4 is the de-Shalit and Talmi result (Eq(33.26) in ref [6]). However, further down the same page they say ``No comparison with experiment can yet be made with mirror nuclei since at least one of each pair is an unstable nucleus''. Note that the main thrust in this section of the current paper is two degrees of separation from this. There has since been abundant data on mirror pairs but we are mainly going to the next level - isoscalar moments of excited states of $N=Z$ nuclei for which the data is much more sparse. Even in later textbooks e.g. Bohr and Mottleson \cite{bohr69}, Lawson \cite{lawson80}, of Talmi \cite{talmi93} there are no discussions of magnetic moments of excited states of $N=Z$ nuclei. The 1993 Talmi book has a very brief discussion of mirror pairs. Bohr and Mottleson show measurements of g factors of excited states for $N \neq Z$ nuclei. These have very large error bars.

Equation (2) holds for $N=Z$ nuclei, in odd-odd as well as even-even, in the single $j$ shell approximation, not only for $T=0$ states, but also for states of any isospin and any total angular momentum $I$. Another interesting general result for an $N=Z$ nucleus in the single-$j$  shell model is that, even with configuration mixing, the expectation value of the isovector magnetic moment in any state is zero. This can be shown by noting that the expression for this expectation value contains an isospin Clebsch-Gordon coefficient (1T00$\mid$T0).  This coefficient vanishes for all integer T.

The Schmidt formula for an isoscalar magnetic moment $g_0$ in the single $j$ shell picture is 
\begin{equation}
g_0 = \left[1 \mp \frac{1}{(2l+1)}\right]g_l \pm \left[\frac{1}{(2l+1)}\right]g_s \qquad \textrm{for} \quad j = l \pm 1/2
\end{equation}

Using $g_l = 0.5$ and $g_s = 0.8796$, as in \cite{zamick77}, this leads to the explicit expressions

\begin{equation}
g_0 = 0.5 + \frac{0.38}{2l+1} \qquad \qquad \textrm{for} \quad j = l+1/2
\end{equation}

\begin{equation}
g_0 = 0.5 - \frac{0.38}{2l+1} \qquad \qquad \textrm{for} \quad j = l-1/2
\end{equation}

It should be noted that for the spin-orbit partners with the same $l$, $j = l + 1/2$ and $j = l - 1/2$, the sum of the expressions (4) and (5) is always 1. Furthermore, in the limit of large $l$, these two expressions converge to $g_0=0.5$, from above and from below, respectively. For example, even for $l$ as small as 3, for $f_{7/2}$ and $f_{5/2}$, each of the two corresponding isoscalar $g$ factors already differ from 0.5 by less than 10\%.

In Table IV below, we use Eq. (4) and (5) to evaluate the isoscalar $g$ factors in the various $(l,j)$ orbitals. The proximity to $g_0=0.5$, and the approach to this value as $l$ increases, are obvious.

\begin{center}
\begin{tabular}{| c | c | c | c |}
\multicolumn{4}{c}{Table IV Isoscalar Schmidt $g$}\\
\multicolumn{4}{c}{factors in the single $j$ shell model} \\ \hline

\multicolumn{2}{| c |}{\emph{j} = \emph{l} + 1/2} & \multicolumn{2}{| c |}{\emph{j} = \emph{l} - 1/2} \\
\multicolumn{2}{| c |}{0.5 +$ \frac{0.38}{2l+1}$} & \multicolumn{2}{| c |}{0.5 - $ \frac{0.38}{2l+1}$} \\ \hline
$s_{1/2}$ & 0.88 & & \\ \hline
$p_{3/2}$ & 0.63 & $p_{1/2}$ & 0.37 \\ \hline
$d_{5/2}$ & 0.58 & $d_{3/2}$ & 0.42 \\ \hline
$f_{7/2}$ & 0.55 & $f_{5/2}$ & 0.45 \\ \hline
$g_{9/2}$ & 0.54 & $g_{7/2}$ & 0.46 \\ \hline
$h_{11/2}$ & 0.53 & $h_{9/2}$ & 0.47 \\ \hline
\end{tabular} \end{center}

In this section it was shown that in many cases for $N=Z$ nuclei, the single-$j$ shell model and the collective model (wherein $g=Z/A=0.5$ for $N=Z$ nuclei) both predict similar values of $g \cong 0.5$ for the isoscalar $g$ factors.  Therefore, one has to be careful in drawing conclusions about the details of the nuclear structure from the proximity to 0.5 of the experimental isoscalar $g$ factor results for $N=Z$ nuclei.  Such is usually not the case for isovector $g$ factors. We also note that isoscalar $g$ factors are of special interest because of their renormalization properties.  For example, it was shown by Mavromatis et al. \cite{mavromatis66}. that for a closed LS shell plus or minus one nucleon, there are no corrections to the magnetic moment in first order perturbation theory and that only the tensor interaction contributes to the renormalization of the isoscalar \emph{g} factors in second order perturbation theory. Also, the renormalization of $g_l$ due to one pion exchange affects only the isovector orbital term - not the isoscalar one.

\section{$g$ Factors of the Ground States of Special Mirror-Nuclei Pairs}

One can apply the results of Table IV, for the isoscalar $g$ factors in a single $j$ shell, to the magnetic moments of the ground states of special mirror-nuclei pairs with closed $LS$ and $jj$ shells plus or minus one nucleon. 
\begin{equation}
\mu _{isoscalar} \equiv \mu _0 = \frac{1}{2} \left(\mu _P + \mu _N \right)
\end{equation}
The isoscalar magnetic moment is $\mu _0 = (\mu (T_z=1/2)+\mu (T_z=-1/2))/2$.

Such an analysis was carried out by Zamick \cite{zamick77} for five special mirror-nuclei pairs corresponding to five different orbitals. In that paper the experimental deviations from the Schmidt values, $\delta \mu _0$, were attributed to an effective operator of the form $g_l \vec{l} + g_s \vec{s} + [Y^2 \sigma]^1$, where the last term comes from an induced $l$-forbidden term. Zamick thus tried to fit the $\delta \mu _0$ data by an expression of the form
\begin{equation}
\langle j \quad |\delta \mu _0| \quad j \rangle = aj \pm \frac{bj}{2l + 1} + c\frac{1 \mp (j + \frac{1}{2})}{2(j + 1)} \qquad \textrm{for } j = l \pm \frac{1}{2}
\end{equation}
Above, $a = \delta g_l$, $b = \delta g_s -\delta g_l$ and $c$ is due to the $[Y^2 \sigma]^1$ terms. The parameters a, b, and c are assumed to be independant of $j$

This fit (see Table II in \cite{zamick77}) led to $\delta g_l = 0.007$, a very small value, $\delta g_s = -0.055$ and $c = 0.050$. Thus, the deviations $\delta \mu _0$ from the Schmidt values could be attributed primarily to changes in $g_s$ and $c$.

It should be noted that the $[Y^2\sigma]^1$ term is not purely phenomological. It can arise from second-order core polarization and meson exchange currents, as was noted, for example, by Towner \cite{towner87} and Shimizu et al.\cite{shimizu74}. These authors calculated the parameters a, b, and c and found them to have a weak dependance on $j$.

To clarify the role of the $[Y^2\sigma]^1$ term, we now investigate in the present paper what would happen in its absence.  We set both $\delta g_l = 0$ (since it is so small) and $c=0$. Then, for each orbit, we vary $g_s$ in expression (2), with $g_l$ fixed at 0.5, to fit the experimental $\mu _0$ for the mirror-nuclei pair. The value of $g_s$ that is thus obtained for each orbit is given in Table V.

\begin{center} Table V - The value \\ 
of $g_s$ obtained from \\ 
fitting the measured \\
$\mu _0$ in the orbit \\
\begin{tabular}{| c | c |} \hline
$s_{1/2}$ & 0.8515 \\ \hline
$p_{1/2}$ & .6920 \\ \hline
$d_{5/2}$ & .8280 \\ \hline
$d_{3/2}$ & .6447 \\ \hline
$f_{7/2}$ & .8313 \\ \hline
\end{tabular} \end{center}

We note that in the absence of the `$l$-forbidden' term, these renormalized values of $g_s$ for the $j=l-1/2$ orbits are consistently smaller than those for the $j=l+1/2$ orbits. The standard value of $g_s = 0.8796$.  Thus the renormalized $g_s$ values for the $j = l + \frac{1}{2}$ orbits are smaller than the standard value by only 6\% or less, but they are smaller by over 20\% for the $j = l- \frac{1}{2}$ orbits. With the inclusion of the '$l$-forbidden' term we can fit all the $\mu _0$ data quite well with the set of the three $j$-independent parameters $g_l$, $g_s$, and $c$. In Table V we compare both the experimental $\mu _0$ and the best-fitted $\mu _0$ to the Schmidt values.

\begin{center} Table VI - Deviations from Schmidt\\
\begin{tabular}{| c | c | c |} \hline
& Exp-Schmidt & Fit-Schmidt  \\ \hline
$s_{1/2}$ & -0.0142 & -0.024 \\ \hline
$p_{1/2}$ & 0.0312 & 0.046 \\ \hline
$d_{5/2}$ & -0.0258 & -0.024 \\ \hline
$d_{3/2}$ & 0.0704 & 0.057 \\ \hline
$f_{7/2}$ & -0.0226 & -0.020 \\ \hline
\end{tabular}\end{center}

It should be noted that for the $j = l+\frac{1}{2}$ orbitals the Exp-Schmidt and the fit-Schmidt results are all negative, while for $j = l-\frac{1}{2}$, they are always positive. Furthermore, the very small deviations of our fitted values from the expreimental ones validates our assumption that a, b, and c in Equation(7) can be treated to a good approximation as state-independent.

In closing, we reaffirm that the robustness of the results that for $N=Z$ even-even and odd-odd nuclei the isoscalar $g$ factors are close to 0.5, requires the fact that not only is one in many cases close to the $LS$ limit but also the fact that for large $A$ the $jj$ limit gives nearly the same answer, $g \cong 0.5$. The $LS$ limit cannot explain why $g$ factors of heavy odd-odd nuclei are close to 0.5, but the combination of $LS$ and $jj$ does.

We thank Gulhan Gurdal and Gerfried Kumbartzki for their help. One of us (S. Y.) thanks the Aresty program at Rutgers University for support.

\bibliographystyle{unsrt}
\bibliography{Sources}

\begin{thebibliography}{10}

\bibitem{talmi71}
I.~Talmi.
\newblock {\em Hyperfine Excited States of Nuclei}.
\newblock Gordon and Breach, New York, 1971.

\bibitem{arima78}
A.~Arima.
\newblock {\em Hyperfine Interactions}, volume~4.
\newblock 1978.

\bibitem{blinstoyle57}
P.~J. Blin-Stoyle.
\newblock {\em Theories of Nuclear Moments}.
\newblock Oxford University Press, 1957.

\bibitem{talmi51}
I.~Talmi.
\newblock {\em Phys. Rev.}, 83:1248, 1951.

\bibitem{stone05}
N.~J. Stone.
\newblock {\em {Atomic Data and Nuclear Data Tables}}, 90:75, 2005.

\bibitem{speidel06}
K.~{-}H. Speidel{,} S. Schielke{,} J. Leske{,} J. Gerber{,} P. Maier-Komor{,}
  S. J. Q. Robinson{,} Y. Y. Sharon{,}~L. Zamick.
\newblock {\em Phys Lett B}, 632:207, 2006.

\bibitem{schielke03}
S.~Schielke{,} K. {-}H. Speidel{,} O. Kenn{,} J. Leske{,} N. Gemein{,} M.
  Offer{,} Y. Y. Sharon{,} L. Zamick{,} J. Gerber{,}~P. Maier-Komor.
\newblock {\em Phys Lett B}, 567:153, 2003.

\bibitem{brown82}
B.~A.~Brown J.
\newblock {\em Phys. G: Nuc. Phys.}, 8:679, 1982.

\bibitem{leske03}
J.~Leske{,} K. {-}H. Speidel{,} O. Kenn{,} S. Schielke{,} G.
  M$\mathrm{\ddot{u}}$ller{,} J. Gerber{,} N. Benczer-Koller{,} G.
  Kumbartzki{,}~P. Maier-Komor.
\newblock {\em Phys Lett B}, 551:249, 2003.

\bibitem{stone08}
N.~J. Stone{,} U. K$\mathrm{\ddot{o}}$ster{,} J. Rikovska Stone{,} D. V.
  Fedorov{,} V. N. Fedoseyev{,} K. T. Flanagan{,} M. Hass{,}~S. Lakshmi.
\newblock {\em Phys Rev C}, 77:067302, 2008.

\bibitem{deshalit63}
A.~de{-}Shalit{,} I.~Talmi.
\newblock {\em {New York Academic Press}}, 1963.

\bibitem{mccullen64}
J.~D. McCullen{,} B.F Bayman{,}~L. Zamick.
\newblock {\em Phys Rev B.}, 134:515, 1964.

\bibitem{bohr69}
A.~Bohr{,} B.~R. Mottleson.
\newblock {\em Nuclear Structure}.
\newblock Benjamin, New York, Vol. I: 1969{,} Vol II: 1975.

\bibitem{lawson80}
R.~D. Lawson.
\newblock {\em Theory of the Nuclear Shell Model}.
\newblock Clarindar Press, Oxford, 1980.

\bibitem{talmi93}
I.~Talmi.
\newblock {\em Simple Models of Complex Nuclei}.
\newblock Harwood Academic Publishers, 1993.

\bibitem{zamick77}
L.~Zamick.
\newblock {\em Phys. Rev.}, C15:824, 1977.

\bibitem{mavromatis66}
H.~A. Mavromatis{,} L. Zamick{,} G.~E. Brown.
\newblock {\em Nuclear Physics}, 80:545, 1966.

\bibitem{towner87}
I.~Towner.
\newblock {\em Physics Reports}, 155:263, 1987.

\bibitem{shimizu74}
K.~Shimizu{,} M. Ichimura{,}~A. Arima.
\newblock {\em Nucl. Phys.}, A226:282, 1974.

\end{thebibliography}
\end{document}